\documentclass[english,preprint]{aastex}
\usepackage[T1]{fontenc}
\usepackage[latin9]{inputenc}
\usepackage{amsbsy}
\usepackage{amssymb}

\makeatletter
\usepackage{amsmath}



\makeatother

\usepackage{babel}
\begin{document}

\title{Dispersion of Magnetic Fields in Molecular Clouds. IV - Analysis
of Interferometry Data}

\author{Martin Houde$^{1,2}$, Charles L. H. Hull$^{3,4}$, Richard L. Plambeck$^{5}$,
John E. Vaillancourt$^{6}$, and Roger H. Hildebrand$^{7,8}$}

\affil{$^{1}$Department of Physics and Astronomy, The University of Western
Ontario, London, ON, N6A 3K7, Canada}

\affil{$^{2}$Division of Physics, Mathematics and Astronomy, California
Institute of Technology, Pasadena, CA 91125, USA }

\affil{$^{3}$Harvard-Smithsonian Center for Astrophysics, 60 Garden St.,
Cambridge, MA, 02138, USA }

\affil{$^{4}$Jansky Fellow of the National Radio Astronomy Observatory,
which is a facility of the National Science Foundation operated under
cooperative agreement by Associated Universities, Inc.}

\affil{$^{5}$Astronomy Department \& Radio Astronomy Laboratory, University
of California, Berkeley, CA 94720-3411, USA }

\affil{$^{6}$Stratospheric Observatory for Infrared Astronomy, Universities
Space Research Association, NASA Ames Research Center, Moffet Field,
CA 94035, USA}

\affil{$^{7}$Department of Astronomy and Astrophysics and Enrico Fermi
Institute, The University of Chicago, Chicago, IL 60637, USA}

\affil{$^{8}$Department of Physics, The University of Chicago, Chicago,
IL 60637, USA}
\begin{abstract}
We expand on the dispersion analysis of polarimetry maps toward applications
to interferometry data. We show how the filtering of low-spatial frequencies
can be accounted for within the idealized Gaussian turbulence model,
initially introduced for single-dish data analysis, to recover reliable
estimates for correlation lengths of magnetized turbulence, as well
as magnetic field strengths (plane-of-the-sky component) using the
Davis-Chandrasekhar-Fermi method. We apply our updated technique to
TADPOL/CARMA data obtained on W3(OH), W3 Main, and DR21(OH). For W3(OH)
our analysis yields a turbulence correlation length $\delta\simeq19$
mpc, a ratio of turbulent-to-total magnetic energy $\left\langle B_{\mathrm{t}}^{2}\right\rangle /\left\langle B^{2}\right\rangle \simeq0.58$,
and a magnetic field strength $B_{0}\sim1.1\:\mathrm{mG}$; for W3
Main $\delta\simeq22$ mpc, $\left\langle B_{\mathrm{t}}^{2}\right\rangle /\left\langle B^{2}\right\rangle \simeq0.74$,
and $B_{0}\sim0.7\:\mathrm{mG}$; while for DR21(OH) $\delta\simeq12$
mpc, $\left\langle B_{\mathrm{t}}^{2}\right\rangle /\left\langle B^{2}\right\rangle \simeq0.70$,
and $B_{0}\sim1.2\:\mathrm{mG}$.
\end{abstract}


\section{Introduction}

Given the difficulties of directly measuring magnetic fields in the
interstellar medium (ISM), with Zeeman observations still being the
only means for achieving this goal \citep{Heiles1997,Crutcher1999,Falgarone2008},
plane of sky linear polarization maps have become in the last few
decades the primary way by which magnetic field studies have been
pushed forward. Qualitative analyses of magnetic fields morphologies
from polarization maps and data have recently been replaced by more
quantitative techniques to provide a better view and understanding
of magnetized turbulence in the ISM. Although structure functions,
developed for the studies of turbulence in general \citep{Frisch1995},
had been previously used with polarization maps to study the large
scale behavior in the orientation of magnetic fields \citep{Kobulnicky1994,Dotson1996}
or of polarized intensities \citep{Beck1999}, more recent works have
introduced novel methods aimed at studying magnetized turbulence on
smaller scales \citep{Diego2008,Hildebrand2009,Houde2009,Houde2011}.

Following the study of \citet{Diego2008} who applied the structure
function of the polarization angle to simulations, \citet{Hildebrand2009}
(hereafter Paper I) applied the technique to actual data obtained
with the Hertz polarimeter \citep{Dowell1998} to generally address
one issue that had been a source of error when such data were used
with the so-called Davis-Chandrasekhar-Fermi (DCF) method \citep{Davis1951,CF1953}.
That is, in Paper I, among other things, a method based on the expected
difference in length scales between the turbulent and ordered (or
large-scale) components of the magnetic field was introduced to remove
the latter's unwanted contribution to the angular dispersion used
in the DCF equation, without having to assume any shape for the ordered
field orientation. A second issue that also affected estimates of
magnetic field strengths with the DCF method is the unavoidable signal
integration across the telescope beam and through the depth of the
sources probed by the observations. This issue, first discussed by
\citet{Myers1991} for the case of a narrow, pencil-like telescope
beam, brings a systematic decrease of the apparent level of turbulence
(or the angular dispersion measured from the polarization pseudo-vectors)
and a corresponding erroneous increase in the field strength obtained
with the DCF equation. \citet{Houde2009} (hereafter Paper II) showed
how this could be properly handled for single-dish observations by
developing an analytical solution for the problem using an isotropic
Gaussian turbulence model. This not only allowed to correct for the
signal integration problem but also made it possible to provide estimates
for magnetized turbulence correlation lengths. For example, from OMC-1
SHARP data \citep{Novak2004,Li2006,Li2008a} they measured a turbulent
correlation length of 16 mpc and a magnetic field strength of approximately
$760\;\mu\mathrm{G}$. The analysis developed in Papers I and II were
since applied in several studies led by different teams of researchers
(see for example, \citealt{Franco2010,Chapman2011,Girart2013,Planck2015}).

Subsequently the dispersion analysis was further developed and applied
to studies of the magnetized turbulent power spectrum and the potential
determination of turbulence dissipation scales (\citealt{Houde2011},
hereafter Paper III), and magnetohydrodynamics (MHD) turbulence anisotropy
in molecular clouds (\citealt{Chitsazzadeh2012}, hereafter Paper
IV). Eventually the Gaussian turbulence model was extended and solved
for the more general case of two-dimensional turbulence and successfully
applied to the synchrotron polarization data of M51 from \citet{Fletcher2011}
to clearly reveal the anisotropy in the turbulent component of the
magnetic field (expected from MHD turbulence theory; \citealt{Goldreich1995})
in this galaxy (\citealt{Houde2013}, hereafter Paper V). More specifically
to this paper, Paper III discussed the issues that arise when the
dispersion analysis is applied to high-resolution interferometry data.
More precisely, the filtering of low spatial frequencies (i.e., extended
structures) inherent to interferometers was shown to render questionable
the application of the Gaussian turbulence model of Paper II for the
dispersion analyses of such polarization data (and to the DCF method,
for example). 

In this paper, we revisit the application of the dispersion analysis
to interferometry data by extending the isotropic Gaussian turbulence
model to account for the low spatial frequency filtering process.
We will specifically focus on the application of the technique and
the ability to obtain reliable quantitative results rather than on
the astrophysical implications of these results for the sources we
will study. We start with a brief summary of the main definitions
and equations for the dispersion analysis in Section \ref{sec:Analysis}
paying special attention to the isotropic Gaussian turbulence model
as developed for single-dish observations in Section \ref{sub:single-dish},
while the generalization to interferometer data is presented in Section
\ref{sub:interferometer}. We then follow with an application of the
new model to TADPOL/CARMA data of W3(OH), W3 Main, and DR21(OH) previously
published by \citet{Hull2014} in Section \ref{sec:Results}, and
we end with a summary and conclusion in Section \ref{sec:Conclusion}.
The details of the data processing and error propagation calculations
will be found in the Appendix at the end.

\section{The Dispersion Analysis \label{sec:Analysis}}

As mentioned earlier, the development of the dispersion analysis,
while taking into account signal integration in the column of gas
subtended by the telescope beam, was initially performed in Paper
II, but the first application to interferometry data with a special
emphasis on the characterization of the magnetized turbulence power
spectrum was done in Paper III. This model was subsequently enhanced
to include anisotropic turbulence in Paper V. For convenience, we
give here a brief summary of the main equations for the isotropic
turbulence case and mainly focus on the Gaussian turbulence approximation
introduced in Paper II for the determination of correlation lengths
$\delta$ and the turbulent-to-total magnetic energy ratio $\left\langle B_{\mathrm{t}}^{2}\right\rangle /\left\langle B^{2}\right\rangle $.
A glossary of the different symbols and parameters appearing in the
equations of this section is given in Table \ref{tab:glossary}.

\begin{deluxetable}{lll}
\tabletypesize{\scriptsize}
\tablewidth{0pt}
\tablecolumns{3}

\tablehead{
\colhead{Parameter} & \colhead{Definition} & \colhead{Reference} 
}

\tablecaption{Glossary
\label{tab:glossary}}

\startdata

$\mathbf{a}$ & Integration variable -- vector position  on the plane of the sky (POS) & Eqs. (\ref{eq:Bbar}),(\ref{eq:autob2}) \\
$a_{2j}$ & Taylor expansion coefficients for $\alpha^{2}\left(\ell\right)$ & Eq. (\ref{eq:model}) \\ 
$\mathbf{B}\left(\mathbf{x}\right)$ & Total three-dimensional magnetic field & Eq. (\ref{eq:Btot}) \\
$\mathbf{B}_{0}\left(\mathbf{x}\right)$ & Three-dimensional ordered magnetic field & Eq. (\ref{eq:Btot}) \\
$\mathbf{B}_{\mathrm{t}}\left(\mathbf{x}\right)$ & Three-dimensional turbulent magnetic field & Eq. (\ref{eq:Btot}) \\
$\overline{\mathbf{B}}\left(\mathbf{r}\right)$ & Total integrated POS magnetic field & Eq. (\ref{eq:Bbar}) \\
$\overline{\mathbf{B}}_{0}\left(\mathbf{r}\right)$ & Integrated POS ordered magnetic field & Eq. (\ref{eq:alpha}) \\
$\overline{\mathbf{B}}_{\mathrm{t}}\left(\mathbf{r}\right)$ & Integrated POS turbulent magnetic field & Eqs. (\ref{eq:b2}),(\ref{eq:autob2}) \\
$\mathbf{e}_{r}$ & Unit vector along $\mathbf{r}$ on the POS & Eq. (\ref{eq:x}) \\
$\mathbf{e}_{z}$ & Unit vector along the line-of-sight & Eq. (\ref{eq:x}) \\
$F_{0}\left(\mathbf{a},z\right)$ & Ordered polarized emission & Eq. (\ref{eq:Bbar}) \\
$H\left(\mathbf{r}\right)$ & Telescope beam profile & Eqs. (\ref{eq:Bbar}),(\ref{eq:autob2}),(\ref{eq:Gaussian_beam}),(\ref{eq:Gaussian_beam_int}) \\
$H\left(\mathbf{k}_{v}\right)$ & Fourier transform of telescope beam profile & Eq. (\ref{eq:b2(kv)_int}) \\
$\boldsymbol{\ell}$ & Distance between measurement pairs (POS) & Sec. \ref{sec:Analysis} \\
$N$ & Number of independent turbulent cells -- interferometer & Eq. (\ref{eq:N_int}) \\
$N_1$ & Number of independent turbulent cells -- single-dish & Eq. (\ref{eq:N1}) \\
$\mathbf{r}$ & Position vector on the POS & Eq. (\ref{eq:x}) \\
$\mathcal{R}_{\mathrm{3\mathrm{D},t}}\left(v,u\right)$ & Autocorrelation of the intrinsic three-dimensional magnetized turbulence & Eq. (\ref{eq:autob2}),(\ref{eq:Gaussian_turb}) \\
$\mathcal{R}_{\mathrm{3\mathrm{D},t}}\left(\mathbf{k}_{v},k_{u}\right)$ & Power spectrum of the intrinsic three-dimensional magnetized turbulence & Eq. (\ref{eq:b2(kv)_int}) \\
$\mathcal{R}_{\mathrm{t}}\left(\mathbf{k}_{v}\right)$ & Power spectrum of integrated two-dimensional magnetized turbulence & Eq. (\ref{eq:b2(kv)_int}) \\
$W_1,W_2$ & Gaussian telescope beam radii (standard deviation equivalent) & Eqs. (\ref{eq:Gaussian_beam}),(\ref{eq:Gaussian_beam_int}) \\
$\mathbf{x}=r\mathbf{e}_{r}+z\mathbf{e}_{z}$ & Three-dimensional position vector & Eq. (\ref{eq:x}) \\
$z$ & Position along the line-of-sight & Eq. (\ref{eq:x}) \\
$\alpha^{2}\left(\ell\right)$ & Normalized autocorrelation of the integrated  ordered POS magnetic field & Eq. (\ref{eq:alpha}) \\
$b^{2}\left(\ell\right)$ & Normalized autocorrelation of the integrated turbulent POS magnetic field & Eqs. (\ref{eq:b2}),(\ref{eq:b2(l)-SD}),(\ref{eq:b2(l)_int}) \\
$b^{2}\left(\mathbf{k}_{v}\right)$ & Power spectrum of the integrated turbulent POS magnetic field & Eq. (\ref{eq:b2(kv)_int}) \\
$\delta$ & Turbulence correlation length & Eq. (\ref{eq:Gaussian_turb}) \\
$\Delta$ & Maximum depth of a molecular cloud along the line-of-sight & Eq. (\ref{eq:Bbar}),(\ref{eq:autob2}) \\
$\Delta^{\prime}$ & Effective depth of a molecular cloud along the line-of-sight & Eqs. (\ref{eq:N1}),(\ref{eq:N2}),(\ref{eq:N12}) \\
$\Delta\Phi\left(\boldsymbol{\ell}\right)$ & Difference in polarization angles between measurement pairs separated by $\boldsymbol{\ell}$ & Sec. \ref{sec:Analysis}, Eq. (\ref{eq:cos}) \\
$\left\langle \cdots\right\rangle $ & Average of some quantity & Eq. (\ref{eq:cos}) \\

\enddata

\end{deluxetable}

Given the difference $\Delta\Phi\left(\boldsymbol{\ell}\right)\equiv\Phi\left(\mathbf{r}\right)-\Phi\left(\mathbf{r}+\boldsymbol{\ell}\right)$
in the polarization angle $\Phi$ measured at two positions separated
by a distance $\boldsymbol{\ell}$ on the plane of the sky, we define
the dispersion function $1-\left\langle \cos\left[\Delta\Phi\left(\ell\right)\right]\right\rangle $
for the signal-integrated magnetic field $\overline{\mathbf{B}}$
with

\begin{equation}
\left\langle \cos\left[\Delta\Phi\left(\ell\right)\right]\right\rangle =\frac{\left\langle \overline{\mathbf{B}}\mathbf{\cdot}\overline{\mathbf{B}}\mathbf{\left(\ell\right)}\right\rangle }{\left\langle \overline{\mathbf{B}}\mathbf{\cdot}\overline{\mathbf{B}}\left(0\right)\right\rangle },\label{eq:cos}
\end{equation}

\noindent where $\left\langle \cdots\right\rangle $ denotes an average,
$\ell=\left|\boldsymbol{\ell}\right|$, and $\left\langle \overline{\mathbf{B}}\mathbf{\cdot}\overline{\mathbf{B}}\mathbf{\left(\ell\right)}\right\rangle \equiv\left\langle \overline{\mathbf{B}}\left(\mathbf{r}\right)\mathbf{\cdot}\overline{\mathbf{B}}\mathbf{\left(\mathbf{r}+\boldsymbol{\ell}\right)}\right\rangle $
is the autocorrelation function of $\overline{\mathbf{B}}$. The signal-integrated
magnetic field is defined with

\begin{equation}
\overline{\mathbf{B}}\left(\mathbf{r}\right)=\iint H\left(\mathbf{r}-\mathbf{a}\right)\left[\frac{1}{\Delta}\int_{0}^{\Delta}F_{0}\left(\mathbf{a},z\right)\mathbf{B}\left(\mathbf{a},z\right)dz\right]d^{2}a,\label{eq:Bbar}
\end{equation}

\noindent where $H\left(\mathbf{r}\right)$ is the beam profile, $\Delta$
is the maximum depth of the cloud along any line of sight, and the
weighting function $F_{0}\left(\mathbf{a},z\right)\geq0$ scales with
the (ordered) polarized emission associated with the magnetic field
$\mathbf{B}\left(\mathbf{a},z\right)$. Whether one chooses the polarized
emission itself for $F_{0}$ or normalizes it beforehand is irrelevant
for the analysis, as the dispersion function is based on a normalized
quantity (i.e., the right-hand side of Equation {[}\ref{eq:cos}{]}).
Any dependency on the amplitude or units of $F_{0}$ is then removed
from the analysis. The position in the cloud is given by

\begin{equation}
\mathbf{x}=r\mathbf{e}_{r}+z\mathbf{e}_{z}\label{eq:x}
\end{equation}

\noindent with $\mathbf{e}_{r}$ and $\mathbf{e}_{z}$ the unit basis
vectors along $\mathbf{r}$ in the plane of the sky and the $z$-axis
along the line of sight, respectively. We decompose the magnetic field
$\mathbf{B\left(\mathbf{x}\right)}$ into an ordered field, $\mathbf{B}_{0}\mathbf{\left(\mathbf{x}\right)}$,
and a turbulent (random), zero-mean component, $\mathbf{B_{\mathrm{t}}\left(\mathbf{x}\right)}$,
with 

\begin{equation}
\mathbf{\mathbf{B\left(\mathbf{x}\right)}}=\mathbf{B}_{0}\mathbf{\mathbf{\mathbf{\left(\mathbf{x}\right)+\mathbf{B}_{\mathrm{t}}\left(\mathbf{x}\right)}}}.\label{eq:Btot}
\end{equation}

\noindent We further assumed stationarity, homogeneity and isotropy
in the magnetic field strength for Equation (\ref{eq:cos}), while
statistical independence between ordered and turbulent components
will also be implied from now on.

Upon inserting Equations (\ref{eq:Bbar}) and (\ref{eq:Btot}) into
Equation (\ref{eq:cos}) it is found that the latter can be expressed
as the sum of turbulent and ordered terms 

\begin{eqnarray}
1-\left\langle \cos\left[\Delta\Phi\left(\ell\right)\right]\right\rangle  & = & \left[b^{2}\left(0\right)-b^{2}\left(\ell\right)\right]+\left[\alpha^{2}\left(0\right)-\alpha^{2}\left(\ell\right)\right]\nonumber \\
 & = & \left\{ b^{2}\left(0\right)+\left[\alpha^{2}\left(0\right)-\alpha^{2}\left(\ell\right)\right]\right\} -b^{2}\left(\ell\right),\label{eq:b_alpha}
\end{eqnarray}

\noindent with the (signal-integrated) ordered and turbulence normalized
autocorrelation functions given by 

\begin{eqnarray}
\alpha^{2}\left(\ell\right) & = & \frac{\left\langle \overline{\mathbf{B}}_{0}\mathbf{\cdot}\overline{\mathbf{B}}_{0}\left(\ell\right)\right\rangle }{\left\langle \overline{\mathbf{B}}\mathbf{\cdot}\overline{\mathbf{B}}\left(0\right)\right\rangle }\label{eq:alpha}\\
b^{2}\left(\ell\right) & = & \frac{\left\langle \overline{\mathbf{B}}_{\mathrm{t}}\mathbf{\cdot}\overline{\mathbf{B}}_{\mathrm{t}}\left(\ell\right)\right\rangle }{\left\langle \overline{\mathbf{B}}\mathbf{\cdot}\overline{\mathbf{B}}\left(0\right)\right\rangle },\label{eq:b2}
\end{eqnarray}

\noindent respectively. As the ordered function $\left[\alpha^{2}\left(0\right)-\alpha^{2}\left(\ell\right)\right]$
is expected to evolve over a much larger spatial scale than $b^{2}\left(\ell\right)$,
we can expand it with a (slowly varying) Taylor series and write

\begin{equation}
b^{2}\left(0\right)+\left[\alpha^{2}\left(0\right)-\alpha^{2}\left(\ell\right)\right]=b^{2}\left(0\right)+\sum_{j=1}^{\infty}a_{2j}\ell^{2j},\label{eq:model}
\end{equation}

\noindent where $b^{2}\left(0\right)$ is simply the turbulent-to-total
magnetic energy ratio (signal-integrated). The difference in scales
between the function given in Equation (\ref{eq:model}) and the signal-integrated
turbulent autocorrelation function $b^{2}\left(\ell\right)$ allows
for their separation, and the subsequent characterization of magnetized
turbulence.

Using Equation (\ref{eq:Bbar}) the autocorrelation of the signal-integrated
turbulent magnetic field can be shown to be

\begin{equation}
\left\langle \overline{\mathbf{B}}_{\mathrm{t}}\mathbf{\cdot}\overline{\mathbf{B}}_{\mathrm{t}}\mathbf{\left(\ell\right)}\right\rangle =\iint\iint H\left(\mathbf{a}\right)H\left(\mathbf{a}^{\prime}+\boldsymbol{\ell}\right)\left[\frac{2}{\Delta}\int_{0}^{\Delta}\left(1-\frac{u}{\Delta}\right)\mathcal{R}_{\mathrm{3\mathrm{D},t}}\left(v,u\right)du\right]d^{2}a^{\prime}d^{2}a,\label{eq:autob2}
\end{equation}

\noindent with $\mathcal{R}_{\mathrm{3\mathrm{D},t}}\left(v,u\right)=\left\langle F_{0}\left(\mathbf{a},z\right)F_{0}\left(\mathbf{a}^{\prime},z^{\prime}\right)\right\rangle \left\langle \mathbf{B}_{\mathrm{t}}\left(\mathbf{a},z\right)\cdot\mathbf{B}_{\mathrm{t}}\left(\mathbf{a}^{\prime},z^{\prime}\right)\right\rangle $,
$u=\left|z^{\prime}-z\right|$, and $v=\left|\mathbf{a}^{\prime}-\mathbf{a}\right|$.
The function $\mathcal{R}_{\mathrm{3\mathrm{D},t}}\left(v,u\right)$
stands for the autocorrelation of the intrinsic magnetized turbulence
(i.e., unaltered by the measurement process). As was discussed in
Paper III, the magnetized turbulence power spectrum is contained in
the Fourier transform of $b^{2}\left(\ell\right)$ (and that of Eq.
{[}\ref{eq:autob2}{]}; see Eq. {[}\ref{eq:b2}{]})

\begin{equation}
b^{2}\left(\mathbf{k}_{v}\right)=\left\Vert H\left(\mathbf{k}_{v}\right)\right\Vert ^{2}\frac{\mathcal{R}_{\mathrm{t}}\left(\mathbf{k}_{v}\right)}{\left\langle \overline{B}^{2}\right\rangle },\label{eq:b2(kv)_int}
\end{equation}

\noindent where $\left\langle \overline{B}^{2}\right\rangle \equiv\left\langle \overline{\mathbf{B}}\mathbf{\cdot}\overline{\mathbf{B}}\left(0\right)\right\rangle $
and $\mathcal{R}_{\mathrm{t}}\left(\mathbf{k}_{v}\right)\equiv\int\mathcal{R}_{\mathrm{3\mathrm{D},t}}\left(\mathbf{k}_{v},k_{u}\right)\mathrm{sinc}^{2}\left(k_{u}\Delta/2\right)dk_{u}$
is the two-dimensional turbulence power spectrum, with $\mathcal{R}_{\mathrm{3\mathrm{D},t}}\left(\mathbf{k}_{v},k_{u}\right)$
the Fourier transform of $\mathcal{R}_{\mathrm{3\mathrm{D},t}}\left(v,u\right)$.
Although we will not be able to achieve this in this paper, data taken
at high enough spatial resolution can reveal the underlying turbulence
power spectrum $\mathcal{R}_{\mathrm{t}}\left(\mathbf{k}_{v}\right)$
by inverting Equation (\ref{eq:b2(kv)_int}) through the removal of
the filtering due to the telescope beam (i.e., $\left\Vert H\left(\mathbf{k}_{v}\right)\right\Vert ^{2}$,
the Fourier transform of the autocorrelated beam).

\subsection{Isotropic Gaussian Turbulence Model -- Single-dish\label{sub:single-dish}}

As was shown in Papers II, IV, and V, using a Gaussian model as an
idealization for isotropic magnetized turbulence leads to an analytical
solution for the dispersion analysis problem when the telescope beam
is also expressed as a Gaussian function. We will thus use the following
expression for the magnetized turbulence autocorrelation function

\begin{equation}
\mathcal{R}_{\mathrm{3\mathrm{D},t}}\left(v,u\right)=\left\langle F_{0}^{2}\right\rangle \left\langle B_{\mathrm{t}}^{2}\right\rangle e^{-\left(v^{2}+u^{2}\right)/2\delta^{2}},\label{eq:Gaussian_turb}
\end{equation}

\noindent with $\delta$ the turbulence correlation length, $\left\langle B_{\mathrm{t}}^{2}\right\rangle =\left\langle \mathbf{B}_{\mathrm{t}}\cdot\mathbf{B}_{\mathrm{t}}\left(0\right)\right\rangle $,
and $\left\langle F_{0}^{2}\right\rangle =\left\langle F_{0}F_{0}\left(0\right)\right\rangle $.
The telescope beam profile of width $W_{1}$ (i.e., its standard deviation
equivalent) is given by

\begin{equation}
H\left(\mathbf{r}\right)=\frac{1}{2\pi W_{1}^{2}}e^{-r^{2}/2W_{1}^{2}}.\label{eq:Gaussian_beam}
\end{equation}

\noindent Given these functions, we find the following solution for
Equation (\ref{eq:b_alpha}) 

\begin{equation}
1-\left\langle \cos\left[\Delta\Phi\left(\ell\right)\right]\right\rangle =\sum_{j=1}^{\infty}a_{2j}\ell^{2j}+\left[\frac{1}{1+N_{1}\left\langle B_{0}^{2}\right\rangle /\left\langle B_{\mathrm{t}}^{2}\right\rangle }\right]\left[1-e^{-\ell^{2}/2\left(\delta^{2}+2W_{1}^{2}\right)}\right],\label{eq:SD-solution}
\end{equation}

\noindent with the number of turbulent cells probed by telescope beam

\begin{equation}
N_{1}=\frac{\left(\delta^{2}+2W_{1}^{2}\right)\Delta^{\prime}}{\sqrt{2\pi}\delta^{3}}\label{eq:N1}
\end{equation}

\noindent and $\Delta^{\prime}$ the effective depth of the region
under study, which can be determined from the autocorrelation function
of the polarized flux (see Sec. 3.2 of Paper II). It follows that
data from a polarization map used to calculate the left-hand side
of Equation (\ref{eq:SD-solution}) can be fitted to the model expressed
on the right-hand side to provide estimates for $\delta$, $\left\langle B_{\mathrm{t}}^{2}\right\rangle /\left\langle B_{0}^{2}\right\rangle $,
and $a_{2j}$. The signal-integrated turbulence autocorrelation function

\begin{equation}
b^{2}\left(\ell\right)=\left[\frac{1}{1+N_{1}\left\langle B_{0}^{2}\right\rangle /\left\langle B_{\mathrm{t}}^{2}\right\rangle }\right]e^{-\ell^{2}/2\left(\delta^{2}+2W_{1}^{2}\right)},\label{eq:b2(l)-SD}
\end{equation}

\noindent is contained in Equation (\ref{eq:SD-solution}). We find
that its width (i.e., $\sqrt{\delta^{2}+2W_{1}^{2}}$) is broadened
through the measurement process by the telescope beam beyond the intrinsic
correlation length $\delta$ of the underlying turbulence. We also
find that the ``true'' relative level of turbulent energy in the
magnetic field is integrated down through averaging among the $N_{1}$
turbulent cells contained in the column of gas with

\begin{eqnarray}
\frac{\left\langle B_{\mathrm{t}}^{2}\right\rangle }{\left\langle B_{0}^{2}\right\rangle } & = & N_{1}\left[\frac{b^{2}\left(0\right)}{1-b^{2}\left(0\right)}\right]\label{eq:ratio-SD}\\
 & \simeq & N_{1}b^{2}\left(0\right),\label{eq:ratio-SD_approx}
\end{eqnarray}

\noindent with the last equation valid when $b^{2}\left(0\right)$$\ll1$.

\begin{figure}
\epsscale{0.9}\plotone{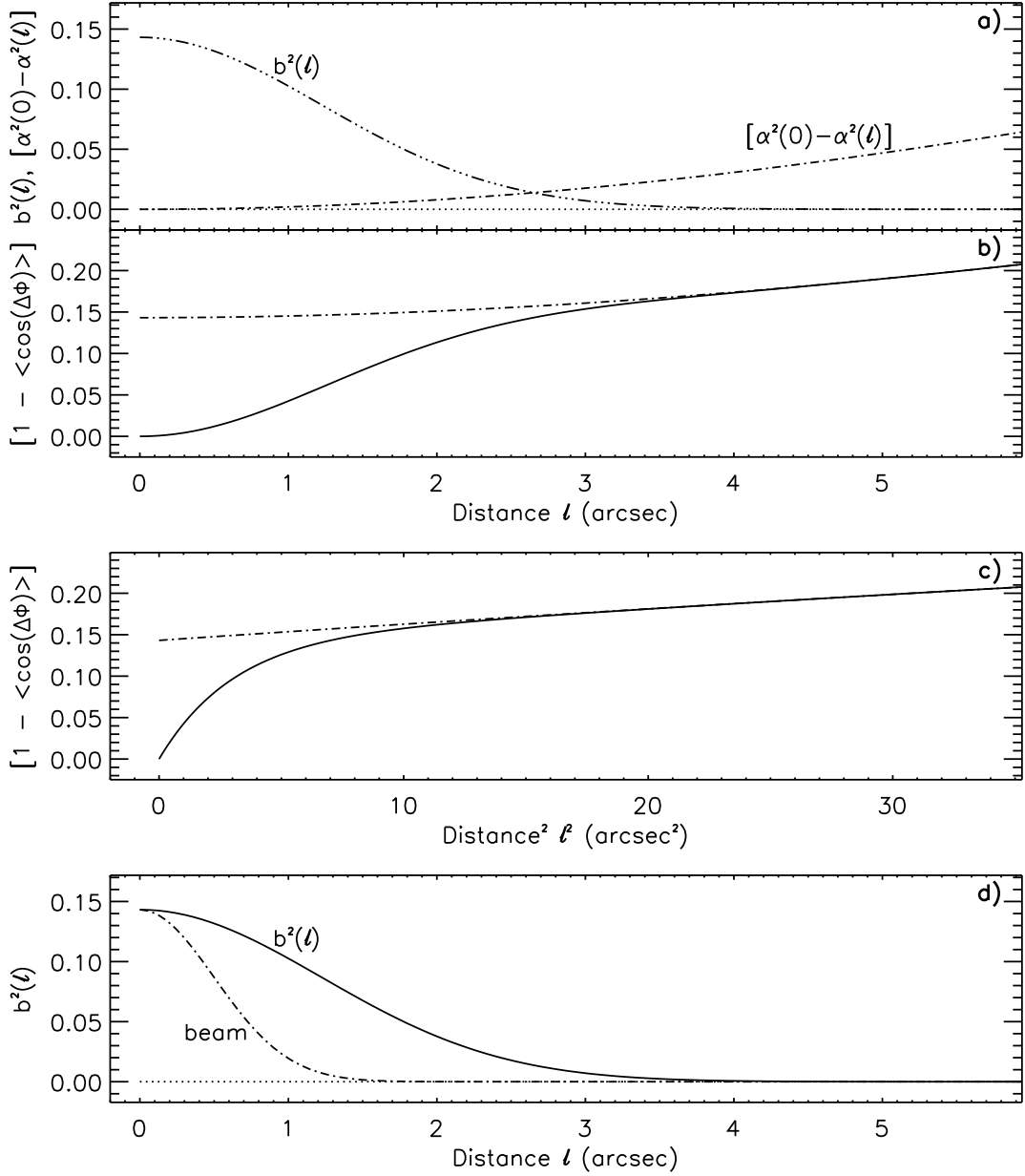}

\caption{\label{fig:Dispersion-SD}Example of an idealized case of Gaussian
turbulence and telescope beam, where we set the turbulence correlation
length $\delta=1\arcsec$, the turbulent-to-ordered magnetic energy
ratio $\left\langle B_{\mathrm{t}}^{2}\right\rangle /\left\langle B_{0}^{2}\right\rangle =0.4$,
the effective depth $\Delta^{\prime}=4\arcsec$, the telescope beam
width $W_{1}=0\farcs5$, and the large-scale coefficients $a_{2}=2\times10^{-3}\:\mathrm{arcsec}^{-2}$
and $a_{4}=-5\times10^{-6}\:\mathrm{arcsec}^{-4}$. \emph{a):} The
turbulence ($b^{2}\left(\ell\right)$; Eq. {[}\ref{eq:b2(l)-SD}{]})
and the ordered ($\alpha^{2}\left(0\right)-\alpha^{2}\left(\ell\right)$)
autocorrelation functions, plotted against $\ell$, combine as in
Equation (\ref{eq:b_alpha}) to yield the dispersion function in \emph{b)}.
The solid and broken-dotted curves in \emph{b) }are for the dispersion
function and $b^{2}\left(0\right)+\left[\alpha^{2}\left(0\right)-\alpha^{2}\left(\ell\right)\right]$,
respectively. For a given data set, the dispersion function is the
starting point for the analysis, i.e., the curves shown in \emph{a)}
are not known a priori. \emph{c):} Same as \emph{b)} but plotted as
a function of $\ell^{2}$ to better show the difference in their length
scales. \emph{d):} The different length scales allow for the separation
of the turbulence autocorrelation function $b^{2}\left(\ell\right)$
(solid curve) from $b^{2}\left(0\right)+\left[\alpha^{2}\left(0\right)-\alpha^{2}\left(\ell\right)\right]$
in \emph{b)} and its recovery. The contribution of the turbulence
correlation length to the broadening of $b^{2}\left(\ell\right)$
is apparent from its excess width in comparison to that of the (autocorrelated)
telescope beam (broken-dotted curve; normalized for convenience).}
\end{figure}

An example of a hypothetical dispersion analysis for the idealized
case of Gaussian turbulence and telescope beam are shown in Figure
\ref{fig:Dispersion-SD}, where we set $\delta=1\arcsec$, $\left\langle B_{\mathrm{t}}^{2}\right\rangle /\left\langle B_{0}^{2}\right\rangle =0.4$,
$\Delta^{\prime}=4\arcsec$, $W_{1}=0\farcs5$, $a_{2}=2\times10^{-3}\:\mathrm{arcsec}^{-2}$,
and $a_{4}=-5\times10^{-6}\:\mathrm{arcsec}^{-4}$. The different
panels show how the turbulence and ordered autocorrelation functions
(in \emph{a)}) combine as in Equation (\ref{eq:b_alpha}) to yield
the dispersion function (in \emph{b)}) obtained from a given data
set. The dispersion function is the starting point for the analysis,
i.e., the curves shown in \emph{a)} are not known a priori. The different
length scales between these two functions allow for the separation
and recovery of the turbulence autocorrelation function $b^{2}\left(\ell\right)$
(solid curve in \emph{d)}) from the ordered component. The contribution
of the turbulence correlation length to the broadening of $b^{2}\left(\ell\right)$
is apparent from its excess width in comparison to that of the (autocorrelated)
telescope beam (broken curve in \emph{d)}).

\subsection{Isotropic Gaussian Turbulence Model -- Interferometry\label{sub:interferometer}}

Equation (\ref{eq:b2(kv)_int}) clearly shows the filtering effect
of the telescope beam on the power spectrum. Of course, this effect
also manifests itself on the profile of the corresponding autocorrelation
function. For the single-dish case shown in Figure \ref{fig:Dispersion-SD},
where the filtering is confined to the high-frequency end of the power
spectrum, the width of the turbulence autocorrelation function $b^{2}\left(\ell\right)$
in panel \emph{d)} has a significant contribution stemming from the
size of the telescope beam. However, we should not expect the appearance
of $b^{2}\left(\ell\right)$ or the dispersion function to be exactly
the same if the measurement was made with an interferometer since,
in this case, the low-frequency end of the spectrum will also be strongly
filtered. Similarly, neither should we expect the analytical solution
to the Gaussian turbulence dispersion problem for interferometry to
be given by Equation (\ref{eq:SD-solution}), which was obtained for
the single-dish case.

\begin{figure}
\plotone{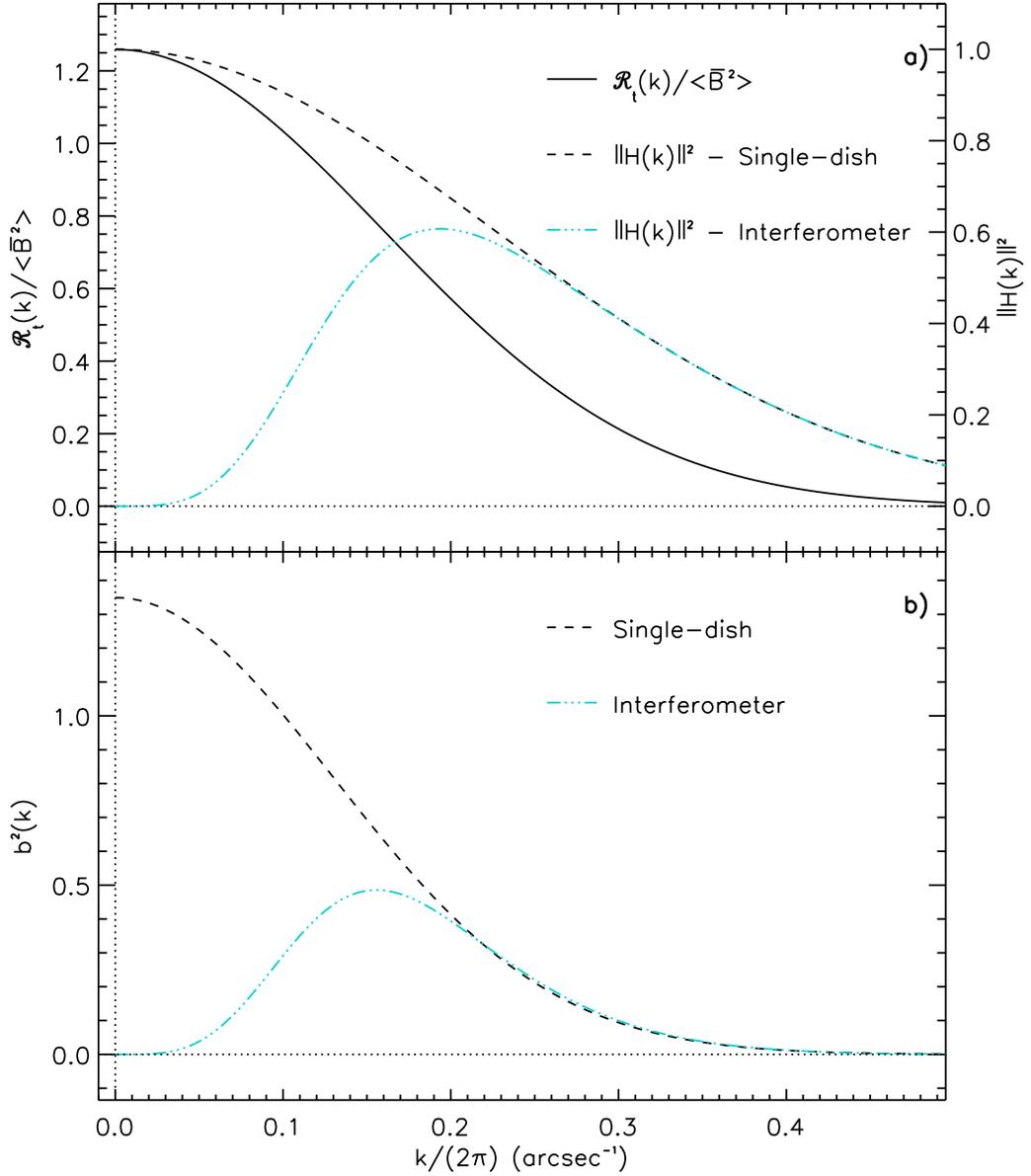}

\caption{\label{fig:Dispersion-spectrum}\emph{Top:} Turbulent power spectrum
$\mathcal{R}_{\mathrm{t}}\left(k\right)/\left\langle \overline{B}^{2}\right\rangle $
that would be observed with a pencil single-dish beam (i.e., with
$W_{1}\rightarrow0$; black solid curve, using the scale on the left),
and the filters corresponding to the single-dish beam for the example
of Figure \ref{fig:Dispersion-SD} (black broken curve) and a corresponding
idealized interferometer beam with $W_{2}=2\arcsec$ (turquoise broken-dotted
curve; both beam profiles use the scale on the right). \emph{Bottom:}
The turbulent power spectra $b^{2}\left(k\right)$ that would be obtained
with the single-dish (black broken curve) and the interferometer (turquoise
broken-dotted curve) beams. }
\end{figure}

Panel \emph{a)} of Figure \ref{fig:Dispersion-spectrum} shows the
turbulent power spectrum $\mathcal{R}_{\mathrm{t}}\left(k\right)/\left\langle \overline{B}^{2}\right\rangle $
that would be observed with a pencil single-dish beam (i.e., with
$W_{1}\rightarrow0$; solid curve, using the scale on the left) for
the example shown in Figure \ref{fig:Dispersion-SD}, as well as the
filter corresponding to the single-dish beam used for these calculations
(black broken curve; right scale). To better display the difference
in the spectral filtering effect, we also show an idealized interferometer
beam spectral profile where the low-frequency component of the single-dish
beam profile was removed by subtracting a Gaussian beam of width $W_{2}=2\arcsec$
(turquoise broken-dotted curve; right scale), to get a better picture
of the effect the so-called dirty beam has on the spectrum. More precisely,
the spatial profile of the interferometer beam was modeled with

\begin{equation}
H\left(\mathbf{r}\right)=\frac{1}{2\pi W_{1}^{2}}e^{-r^{2}/2W_{1}^{2}}-\frac{1}{2\pi W_{2}^{2}}e^{-r^{2}/2W_{2}^{2}},\label{eq:Gaussian_beam_int}
\end{equation}

\noindent with $W_{1}=0\farcs5$, as previously stated, and $W_{2}=2\arcsec$.
We again note that $W_{1}$ and $W_{2}$ are for the standard deviation
equivalent of the corresponding Gaussian beams. This twin-Gaussian
profile representation of an interferometer beam is the one we will
use for the rest of the analyses presented in this paper. It is, in
a sense, an extension of the usual single-Gaussian profile commonly
used for synthesized beams in interferometry, but it has the advantage
of more accurately modeling the spatial filtering of extended structures
caused by the beam. 

Using this twin-Gaussian model for the interferometer beam, it becomes
straightforward to generalize the analytical single-dish solution
for Gaussian turbulence, given by Equations (\ref{eq:SD-solution})
and (\ref{eq:b2(l)-SD}), by substituting Equation (\ref{eq:Gaussian_beam_int})
in the place of Equation (\ref{eq:Gaussian_beam}) into Equation (\ref{eq:autob2}).
Panel \emph{b)} of Figure \ref{fig:Dispersion-spectrum} shows the
resulting integrated power spectrum $b^{2}\left(k\right)$ for each
beam, calculated using Equation (\ref{eq:b2(kv)_int}). The difference
in filtering between the two kinds of measurements is made clear.

The interferometer solution to the idealized Gaussian turbulence problem
is given by

\begin{eqnarray}
1-\left\langle \cos\left[\Delta\Phi\left(\ell\right)\right]\right\rangle  & = & \sum_{j=1}^{\infty}a_{2j}\ell^{2j}+\left[\frac{N}{1+N\left\langle B_{0}^{2}\right\rangle /\left\langle B_{\mathrm{t}}^{2}\right\rangle }\right]\left\{ \frac{1}{N_{1}}\left[1-e^{-\ell^{2}/2\left(\delta^{2}+2W_{1}^{2}\right)}\right]\right.\nonumber \\
 &  & \left.\frac{1}{N_{2}}\left[1-e^{-\ell^{2}/2\left(\delta^{2}+2W_{2}^{2}\right)}\right]-\frac{2}{N_{12}}\left[1-e^{-\ell^{2}/2\left(\delta^{2}+W_{1}^{2}+W_{2}^{2}\right)}\right]\right\} ,\label{eq:int_solution}
\end{eqnarray}

\noindent with $N_{1}$ still given by Equation (\ref{eq:N1}) and

\begin{eqnarray}
N_{2} & = & \frac{\left(\delta^{2}+2W_{2}^{2}\right)\Delta^{\prime}}{\sqrt{2\pi}\delta^{3}}\label{eq:N2}\\
N_{12} & = & \frac{\left(\delta^{2}+W_{1}^{2}+W_{2}^{2}\right)\Delta^{\prime}}{\sqrt{2\pi}\delta^{3}}\label{eq:N12}\\
N & = & \left(\frac{1}{N_{1}}+\frac{1}{N_{2}}-\frac{2}{N_{12}}\right)^{-1}.\label{eq:N_int}
\end{eqnarray}

\noindent The corresponding normalized signal-integrated turbulence
autocorrelation function is 

\begin{equation}
b^{2}\left(\ell\right)=\left[\frac{N}{1+N\left\langle B_{0}^{2}\right\rangle /\left\langle B_{\mathrm{t}}^{2}\right\rangle }\right]\left[\frac{1}{N_{1}}e^{-\ell^{2}/2\left(\delta^{2}+2W_{1}^{2}\right)}+\frac{1}{N_{2}}e^{-\ell^{2}/2\left(\delta^{2}+2W_{2}^{2}\right)}-\frac{2}{N_{12}}e^{-\ell^{2}/2\left(\delta^{2}+W_{1}^{2}+W_{2}^{2}\right)}\right].\label{eq:b2(l)_int}
\end{equation}

\begin{figure}
\plotone{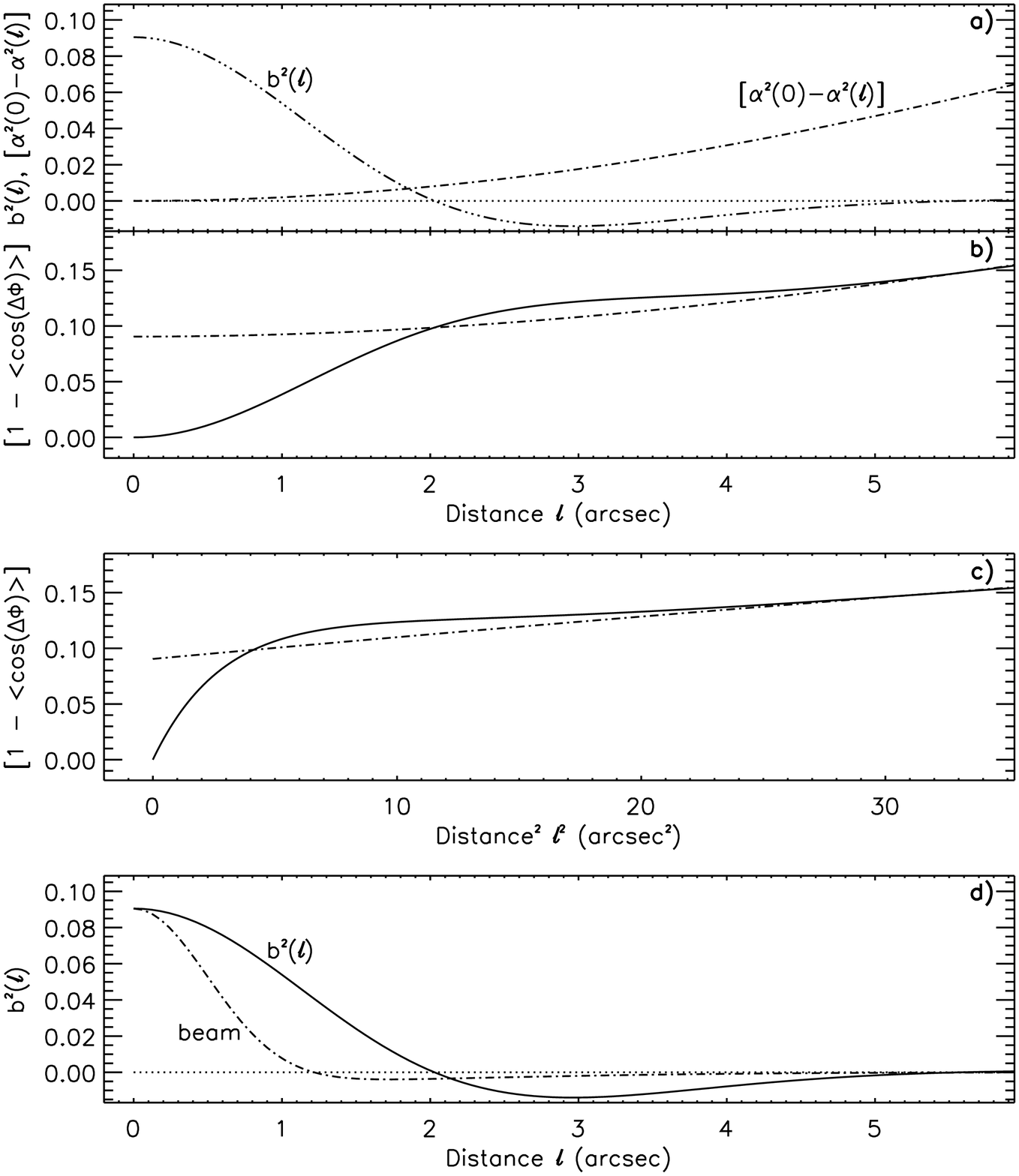}

\caption{\label{fig:Dispersion-int}Same as Figure \ref{fig:Dispersion-SD}
but for an interferometer with $W_{2}=2\arcsec$. The large-scale
component is once again shown with the broken-dotted curve in panels
\emph{b)} and \emph{c)}.}
\end{figure}

\noindent For a given polarimetry map, estimates for $\delta$, $\left\langle B_{\mathrm{t}}^{2}\right\rangle /\left\langle B_{0}^{2}\right\rangle $,
and $a_{2j}$ are obtained by fitting the right-hand side of Equation
(\ref{eq:int_solution}) to the data (on the left-hand side). It is
also straightforward to verify that this interferometry solution tends
to the single-dish solution when $W_{2}\rightarrow\infty$, as would
be expected. Figure \ref{fig:Dispersion-int} shows the corresponding
dispersion function resulting from the interferometer case of the
Gaussian turbulence example used in Figure \ref{fig:Dispersion-SD}.
We have kept the ordered component the same to facilitate the comparison
between the single-dish and interferometer cases. We note the difference
in the appearance of the large-scale component in relation to the
dispersion function in panels \emph{b)} and \emph{c)}, which brings
an oscillatory behavior in the autocorrelation function $b^{2}\left(\ell\right)$
in panel \emph{d)}. These oscillations ensure that

\begin{eqnarray}
2\pi\int_{0}^{\infty}b^{2}\left(\ell\right)\ell d\ell & = & b^{2}\left(\mathbf{k}_{v}=0\right)\nonumber \\
 & = & 0,\label{eq:zero_area}
\end{eqnarray}

\noindent as required for such an interferometer beam. We finally
note that the turbulent-to-total energy ratio is still given by Equation
(\ref{eq:ratio-SD}), but with $N_{1}$ replaced by $N$.

\section{Results -- Analysis of CARMA Data\label{sec:Results}}

We now apply our Gaussian turbulence analysis to interferometry data
on the W3(OH), W3 Main, and DR21(OH) molecular clouds obtained with
CARMA at a frequency of 223.821 GHz. These data were presented and
discussed in detail in \citet{Hull2014}, where more information will
be found. In all cases, we used data points where $p\geq2\sigma_{p}$,
with $p$ and $\sigma_{p}$ the polarization level and its uncertainty,
respectively, and $I\geq3\sigma_{I}$, with $I$ and $\sigma_{I}$
the Stokes $I$ intensity and its uncertainty, respectively. The geometric
mean of the full-width-half-magnitude (FWHM) of the synthesized beam
(divided by $\sqrt{8\ln\left(2\right)}$) was used for $W_{1}$, while
the value for $W_{2}$ was chosen to account for the low-frequency
response and filtering of the telescope dirty beam (see below). The
value $\Delta^{\prime}$ was determined by calculating the (half-)width
of the autocorrelation function of the polarized flux at half of the
maximum amplitude. The dispersion functions were then calculated from
the polarization data using the left-hand side of Equation (\ref{eq:int_solution})
and the Gaussian model fitted for $\delta$, $\left\langle B_{\mathrm{t}}^{2}\right\rangle /\left\langle B_{0}^{2}\right\rangle $,
and $a_{2j}$ on the right-hand side of that same equation. Although
our aforementioned selection criteria for $\sigma_{p}$ and $\sigma_{I}$
reduce the impact of measurement uncertainties on the analysis, we
note that the dispersion function is corrected for corresponding biases
(not unlike the way the polarization level is usually corrected in
linear polarization data). More details for these calculations, and
the propagation of errors, will be found in the Appendix at the end. 

In order to provide an estimate of the magnetic field strength for
each source, we measured that total flux $S_{\nu}$ on the corresponding
map and converted it to a total mass 

\begin{equation}
M_{\mathrm{gas}}=\frac{S_{\nu}d^{2}}{\kappa_{\nu}B_{\nu}\left(T_{\mathrm{d}}\right)},\label{eq:M_gas}
\end{equation}

\noindent where $d$ is the distance to the source, $T_{\mathrm{d}}$
is the dust temperature, and $\kappa_{\nu}$ and $B_{\nu}\left(T_{\mathrm{d}}\right)$
are, respectively, the enhanced mass absorption cross section and
the Planck function at the frequency of the observations \citep{Chini1997}.
We then estimated the approximate size of the source on the sky to
determine its volume and mean mass density, the latter being converted
to a number density by assuming a mean molecular mass of $2.3$. The
mean mass density estimate $\rho$ was then used with suitable line
width information for the one-dimensional turbulence velocity dispersion
$\sigma\left(v\right)$, and the value for $\left\langle B_{\mathrm{t}}^{2}\right\rangle /\left\langle B^{2}\right\rangle $
from our dispersion analysis to evaluate the strength of the plane
of the sky component of the magnetic field with the DCF method \citep{Davis1951, CF1953}

\begin{equation}
B_{0}\simeq\sqrt{4\pi\rho}\,\sigma\left(v\right)\left[\frac{\left\langle B_{\mathrm{t}}^{2}\right\rangle }{\left\langle B^{2}\right\rangle }\right]^{-1/2}.\label{eq:DCF}
\end{equation}

\noindent Finally, although there are other suitable candidates for
a dispersion analysis in the TADPOL sample of \citet{Hull2014}, W3(OH),
W3 Main, and DR21(OH) were chosen because of the large number of independent
polarization measurements available for these sources (resulting in
better statistics) and their relative closeness. We present the results
for each sources below, and provide a summary in Table \ref{tab:results}.

\subsection{W3(OH)}

\begin{figure}
\plotone{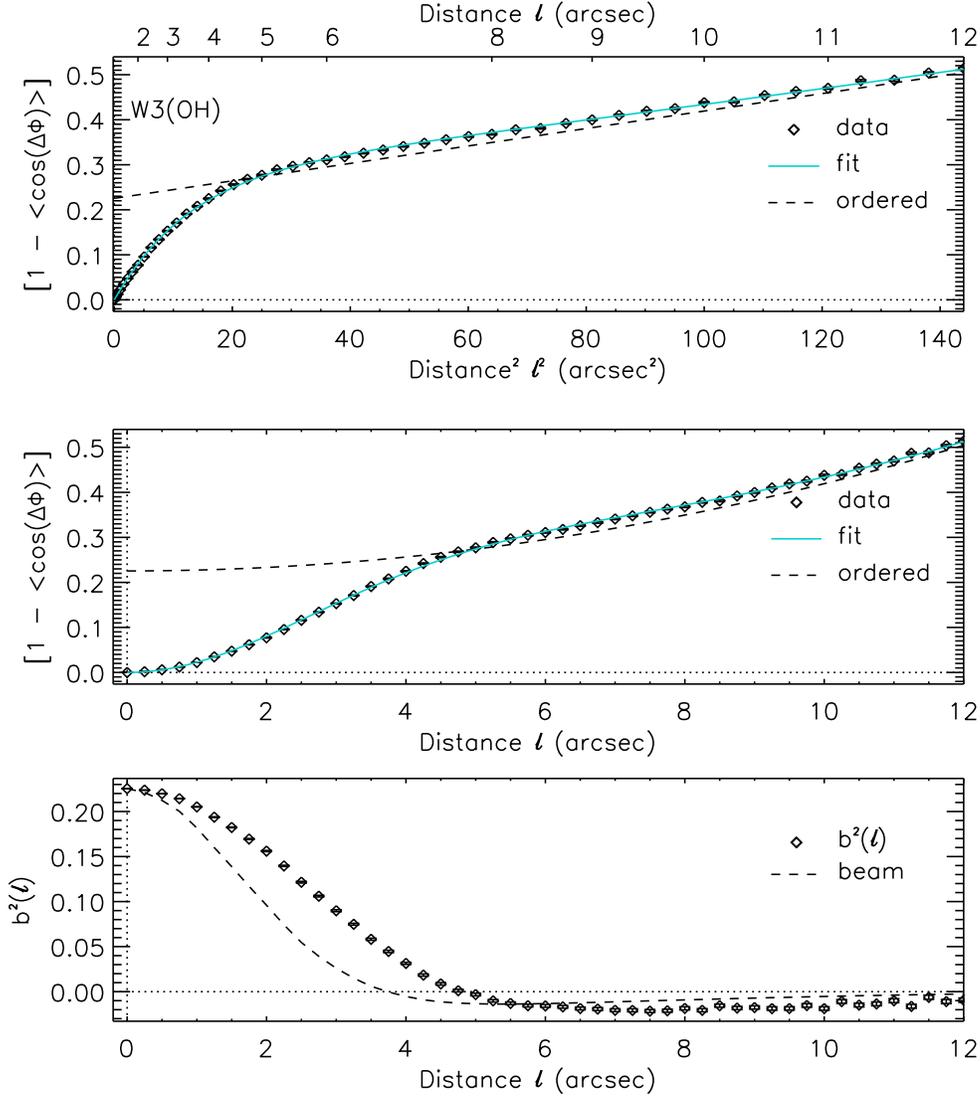}

\caption{\label{fig:W3(OH)-dispersion}Dispersion analysis of the CARMA W3(OH)
data. \emph{Top:} The dispersion function $1-\left\langle \cos\left[\Delta\Phi\left(\ell\right)\right]\right\rangle $
of the data (symbols) plotted as a function of $\ell^{2}$ to better
show the difference in scale for the turbulent and ordered components.
The ordered component (see Equation {[}\ref{eq:model}{]}) is also
shown with the broken curve. The least-squares fit of the Gaussian
turbulence model given in Equation (\ref{eq:int_solution}) is plotted
in turquoise (solid curve). \emph{Middle:} Same as the top panel but
as a function of $\ell$. \emph{Bottom:} The resulting signal-integrated
turbulence autocorrelation function $b^{2}\left(\ell\right)$ (symbols)
is seen to exhibit an excess in its width relative to that of the
autocorrelated beam (broken curve; the beam function is given by Equation
{[}\ref{eq:Gaussian_beam_int}{]}). The fit to the data yields a turbulence
correlation length $\delta\simeq1\farcs92\pm0\farcs02$ (or $19.0\pm0.2$
mpc at the distance of W3(OH)) and a turbulent-to-total magnetic energy
ratio $\left\langle B_{\mathrm{t}}^{2}\right\rangle /\left\langle B^{2}\right\rangle \simeq0.58\pm0.01$.}
\end{figure}

W3(OH) is an active high-mass star-forming region located some 2040
pc away at $\mathrm{RA}(\mathrm{J2000})=2^{\mathrm{h}}27^{\mathrm{m}}03\fs9$,
$\mathrm{Decl}(\mathrm{J2000})=-61^{\circ}52\arcmin24\farcs6$ \citep{Hachisuka2006}.
The polarization map on which our analysis was performed can be found
in Figure 5 of \citet{Hull2014}. The FWHM of the synthesized telescope
beam for these observations is $2\farcs8\times2\farcs6$ at a $\mathrm{PA}=12.4^{\circ}$.
The beam sizes (i.e., their standard deviation equivalent) used for
the twin-Gaussian profile are $W_{1}=1\farcs2$ and $W_{2}=6\farcs6$,
while $\Delta^{\prime}=10\farcs5$ as determined from the autocorrelation
function of the polarized flux. The results of the dispersion analysis
for this source are shown in Figure \ref{fig:W3(OH)-dispersion},
where the top panel is a plot of the dispersion function $1-\left\langle \cos\left[\Delta\Phi\left(\ell\right)\right]\right\rangle $
of the data (symbols) as a function of $\ell^{2}$ to better show
the difference in scale between the turbulent and ordered components.
The center panel of the figure reveals the same information, but this
time with the dispersion function plotted as a function of $\ell$.
The bottom panel yields the resulting signal-integrated turbulence
autocorrelation function $b^{2}\left(\ell\right)$ (symbols), which
is seen to exhibit an excess in its width relative to that of the
autocorrelated beam (broken curve; the beam function is given by Equation
{[}\ref{eq:Gaussian_beam_int}{]}); this is a signature of the intrinsic
magnetized turbulence present in the medium under study. The fit to
the data yields a magnetized turbulence correlation length $\delta=1\farcs92\pm0\farcs02$
(or $19.0\pm0.2$ mpc at the distance of W3(OH)) and a turbulent-to-total
magnetic energy ratio $\left\langle B_{\mathrm{t}}^{2}\right\rangle /\left\langle B^{2}\right\rangle =0.58\pm0.01$.
It is therefore apparent that a significant fraction of the magnetic
energy is in the form of turbulence. The number of turbulent cells
$N$ contained in the column of gas subtended by the telescope beam
was found to be $N=4.67\pm0.04$. Although the fit to the data is
good, it is important to note that the values obtained for $\delta$,
$\left\langle B_{\mathrm{t}}^{2}\right\rangle /\left\langle B^{2}\right\rangle $,
and $N$ are only valid within the framework of the Gaussian turbulence
model. This is an idealization that is certainly not realized for
molecular clouds and the ISM in general, i.e., turbulence is not Gaussian
in nature. Furthermore, we do not precisely know the value of the
effective depth of clouds $\Delta^{\prime}$, and this can have a
significant effect on the uncertainties derived in the analysis. For
example, the estimated values for the turbulent-to-ordered magnetic
energy ratio $\left\langle B_{\mathrm{t}}^{2}\right\rangle /\left\langle B_{0}^{2}\right\rangle $
and the number of turbulent cells $N$, and their uncertainties, scale
linearly with $\Delta^{\prime}$.

\begin{figure}
\plotone{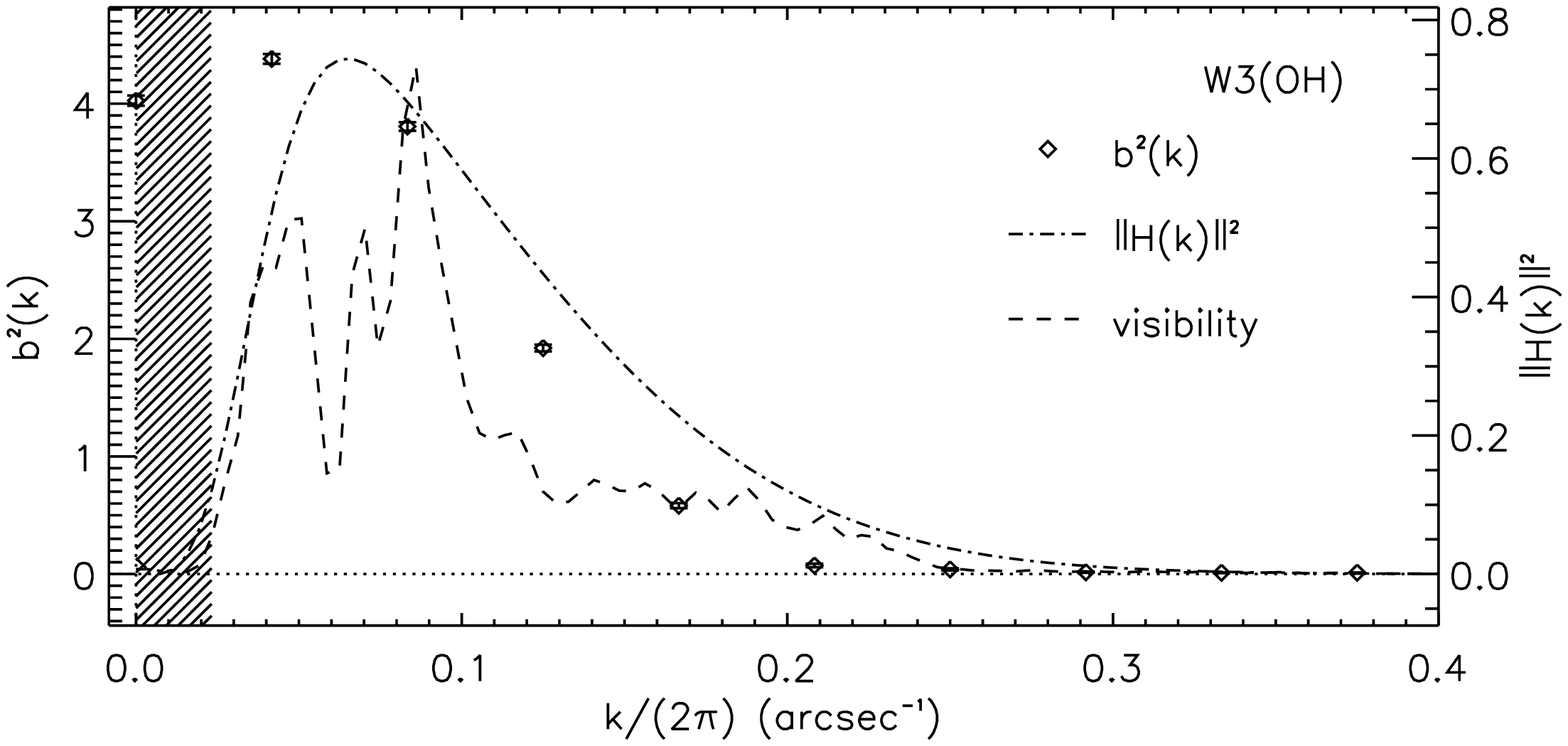}

\caption{\label{fig:W3(OH)-spectrum}The signal-integrated turbulence power
spectrum $b^{2}\left(k\right)$ (symbols) obtained through the Fourier
transform of $b^{2}\left(\ell\right)$ in the bottom panel of Figure
\ref{fig:W3(OH)-dispersion} for W3(OH). The spectral shape of the
autocorrelated dirty beam is also shown (broken curve; ``visibility'')
to better visualize the spectral filtering imposed on the data by
the interferometer. The synthesized beam for the data has $W_{1}=1\farcs2$,
while we subtracted another Gaussian beam component with $W_{2}=6\farcs6$
to model the low-frequency filtering. The resulting autocorrelated
twin-Gaussian beam has the spectral shape shown by the broken-dotted
curve ($\left\Vert H\left(k\right)\right\Vert ^{2}$). The shaded
part of the spectrum at low frequencies outlines the approximate region
where data cannot be trusted in view of the filtering due to the telescope
dirty beam. For this analysis only the datum at $k=0$ is affected
(see text). }
\end{figure}

Figure \ref{fig:W3(OH)-spectrum} shows the signal-integrated turbulence
power spectrum $b^{2}\left(k\right)$ ($k=\left|\mathbf{k}_{v}\right|$;
symbols) obtained through the Fourier transform of $b^{2}\left(\ell\right)$,
taken from the bottom panel of Figure \ref{fig:W3(OH)-dispersion}.
The spectral shape of the autocorrelated dirty beam is also shown
(broken curve; ``visibility'') to better visualize the spectral
filtering imposed on the data by the interferometer. According to
our earlier statement, the synthesized beam for the data has $W_{1}=1\farcs2$,
while we subtracted another Gaussian beam component with $W_{2}=6\farcs6$
to model the low-frequency filtering. The resulting autocorrelated
twin-Gaussian beam has the spectral shape shown by the broken-dotted
curve ($\left\Vert H\left(k\right)\right\Vert ^{2}$) and corresponds
to the Fourier transform of the autocorrelated beam shown in the bottom
panel of Figure \ref{fig:W3(OH)-dispersion}. We note that the power
spectrum $b^{2}\left(k\right)$ does not go to zero at $k=0$, as
would be expected from Equation (\ref{eq:zero_area}). This is likely
due to the fact that the dispersion function cannot be evaluated at
sufficiently large enough values for $\ell$ (i.e., it is truncated),
which causes aliasing in the power spectrum near $k=0$, although
it is also possible that the CLEANing process of the interferometry
data could add signals at low frequencies in the power spectrum \citep{Thompson2004}.
Whatever the case, it follows that the value of $b^{2}\left(k=0\right)$
is erroneous and should not be trusted. The same can be said for the
shaded part of the spectrum at low frequencies, which outlines the
approximate region that is heavily filtered by the telescope dirty
beam. 

We can provide an approximate value for the magnetic field strength
(plane-of-the-sky component) from the fit parameters using the DCF
method \citep{Davis1951, CF1953} with Equation (\ref{eq:DCF}). The
total flux measured from the CARMA map is $S_{\nu}=5.7\:\mathrm{Jy}$,
which from Equation (\ref{eq:M_gas}) translates to a total mass of
$M_{\mathrm{gas}}=149\:M_{\sun}$ for this object with $T_{\mathrm{d}}=30\:\mathrm{K}$
and $\kappa_{\nu}\simeq0.02\;\mathrm{cm^{2}\:g^{-1}}$ \citep{Chini1997}.
This mass is approximately contained within an ellipse of $\simeq10\arcsec\times6\arcsec$
in size (FWHM), yielding a mean density of $\simeq1.3\times10^{6}\;\mathrm{cm^{-3}}$
(or $\rho\simeq5.0\times10^{-18}\:\mathrm{g}\,\mathrm{cm}^{-3}$).
For a measure of $\sigma\left(v\right)$ we follow the prescription
given in Paper II and use the line width from a suitable ion of similar
effective density as that at which the dust emission is detected.
We thus find the line width of $2.8\:\mathrm{km\:s}^{-1}$ from our
own $\mathrm{HCO}{}^{+}\left(4\rightarrow3\right)$ (unpublished)
measurements obtained at the Caltech Submillimeter Observatory (CSO)
at the peak flux position in W3(OH). When scaling down this value
to account for the larger CSO beam (FWHM of $\simeq18\arcsec$) using
a $\sim L^{0.5}$ scaling law for the velocity dispersion (where $L$
is the length-scale), we have $\sigma\left(v\right)\sim1.1\:\mathrm{km}\:\mathrm{s}^{-1}$,
and we obtain $B_{0}\sim1.1\:\mathrm{mG}$ from Equation (\ref{eq:DCF}).
Errors in $B_{0}$ are very difficult to quantify. Since it is subject
to significant uncertainties in $\Delta^{\prime}$, $\rho$, and $\sigma\left(v\right)$,
as well as those intrinsic to the DCF method, we suggest that our
estimate for $B_{0}$ determined this way is accurate to a factor
of about three. Nonetheless, this value is consistent with the Zeeman
$\mathrm{CN}\left(1\rightarrow0\right)$ measurement of \citet{Falgarone2008}
who also found a value of $\sim1.1\:\mathrm{mG}$ for the line of
sight component of the magnetic field in this source.

\subsection{W3 Main}

\begin{figure}
\plotone{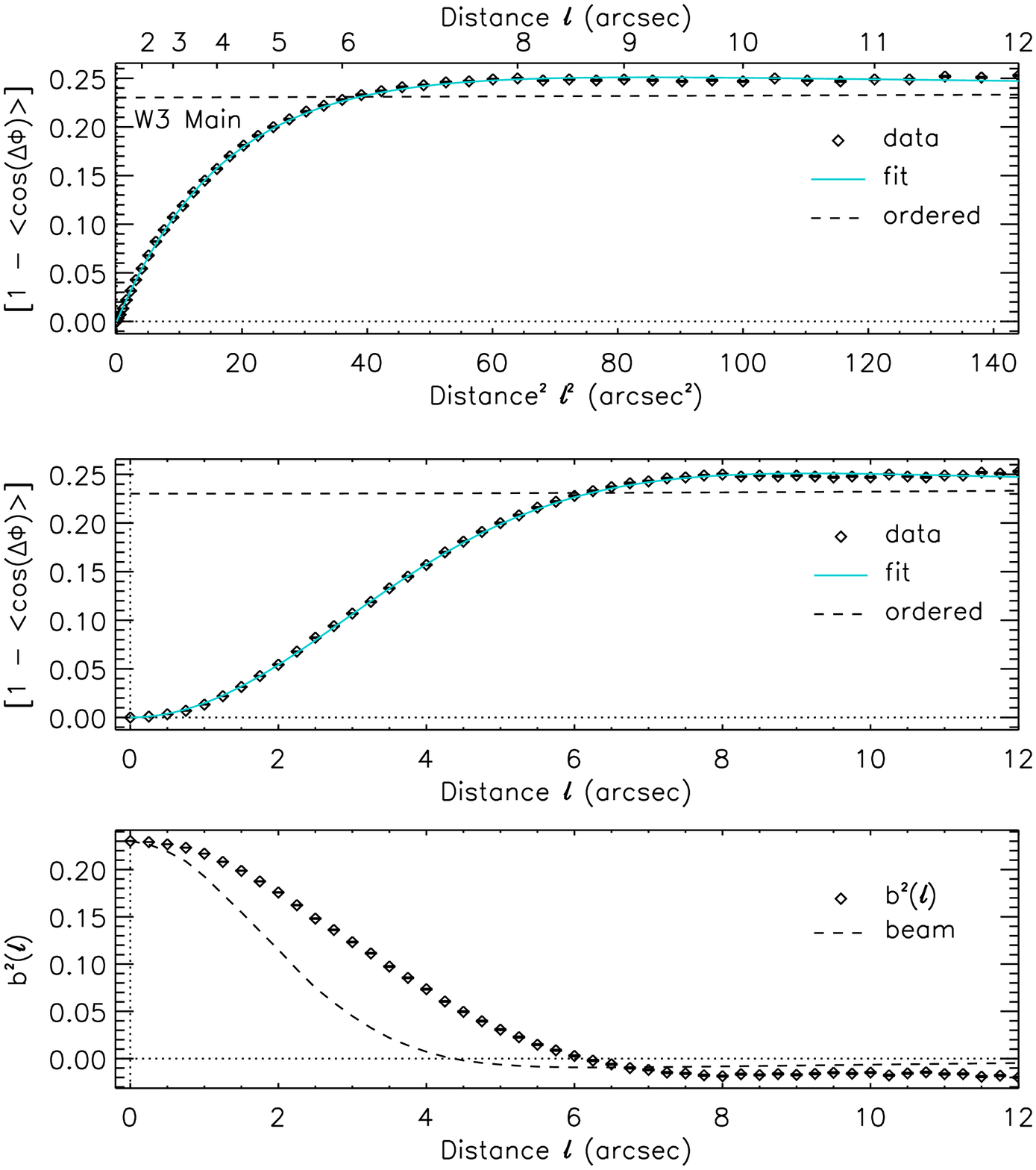}

\caption{\label{fig:W3Main-dispersion}Same as Figure \ref{fig:W3(OH)-dispersion}
but for W3 Main, where the fit to the data yields $\delta\simeq2\farcs35\pm0\farcs03$
(or $22.2\pm0.3$ mpc at the distance of W3 Main) and $\left\langle B_{\mathrm{t}}^{2}\right\rangle /\left\langle B^{2}\right\rangle \simeq0.74\pm0.01$.}
\end{figure}

\begin{figure}
\plotone{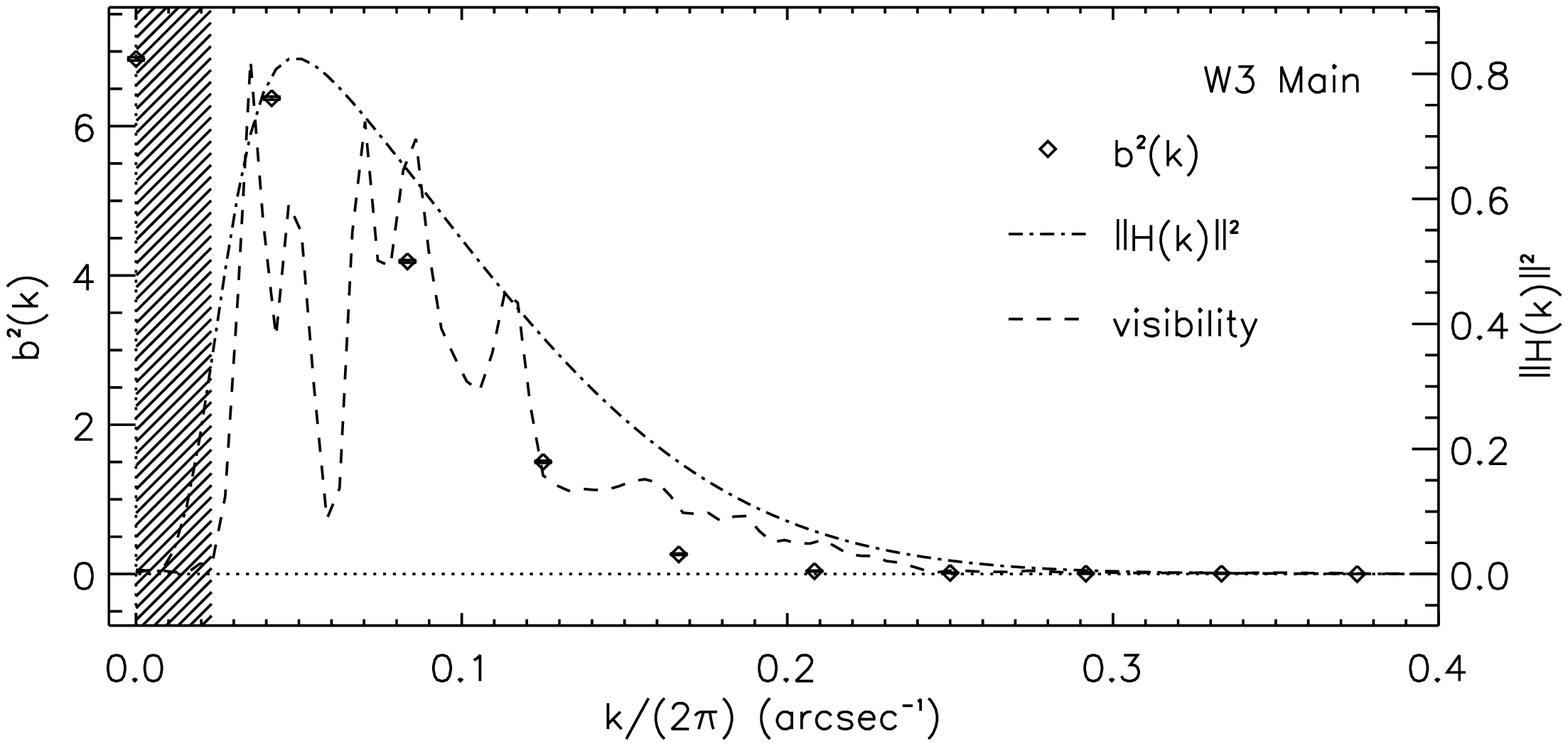}

\caption{\label{fig:W3Main-spectrum}Same as Figure \ref{fig:W3(OH)-spectrum}
but for W3 Main, where $W_{1}=1\farcs2$ and $W_{2}=9\farcs3$.}
\end{figure}

W3 Main is a massive star-forming region located some 1950 pc away
in the outer region of the Galaxy. The CARMA map on which our analysis
is performed is centered on the position of W3 IRS5 at $\mathrm{RA}(\mathrm{J2000})=2^{\mathrm{h}}25^{\mathrm{m}}40\fs6$,
$\mathrm{Decl}(\mathrm{J2000})=-62^{\circ}05\arcmin51\farcs6$, and
will be found in Figure 4 of \citet{Hull2014}. The FWHM of the synthesized
telescope beam for these observations is $3\farcs0\times2\farcs9$
at a $\mathrm{PA}=-27.0^{\circ}$, and the beam sizes used for the
twin-Gaussian profile in our analysis are $W_{1}=1\farcs2$ and $W_{2}=9\farcs3$,
while $\Delta^{\prime}=31\farcs3$. The results of the dispersion
analysis for this source are presented in Figure \ref{fig:W3Main-dispersion}
and Figure \ref{fig:W3Main-spectrum}. As for W3(OH), our model yielded
a good fit to the data with $\delta=2\farcs35\pm0\farcs03$ (or $22.2\pm0.3$
mpc at the distance of W3 Main), $\left\langle B_{\mathrm{t}}^{2}\right\rangle /\left\langle B^{2}\right\rangle =0.74\pm0.01$,
and $N=9.58\pm0.04$. 

The total flux measured from the CARMA map is $3.5\:\mathrm{Jy}$,
yielding a total mass of $83\:M_{\sun}$ (we again use a dust temperature
of $30\:\mathrm{K}$ and $\kappa_{\nu}\simeq0.02\;\mathrm{cm^{2}\:g^{-1}}$)
approximately contained within two circles of $9.8\arcsec$ and $15.6\arcsec$
(FWHM), respectively, as seen on the plane of the sky, which imply
a mean density of $\simeq4.7\times10^{5}\:\mathrm{cm^{-3}}$ (or $\rho\simeq1.8\times10^{-18}\:\mathrm{g}\,\mathrm{cm}^{-3}$).
Using a velocity dispersion of $3.0\;\mathrm{km\:s}^{-1}$ from the
CSO $\mathrm{HCO}{}^{+}\left(4\rightarrow3\right)$ observations of
\citet{Houde2000b}, employing the same power law scaling as before,
we find $\sigma\left(v\right)\sim1.2\:\mathrm{km}\:\mathrm{s}^{-1}$
and $B_{0}\sim0.7\:\mathrm{mG}$. Presumably precise within a factor
of three, we find a magnetic field strength close to that calculated
for W3(OH).

\subsection{DR21(OH)}

\begin{figure}
\plotone{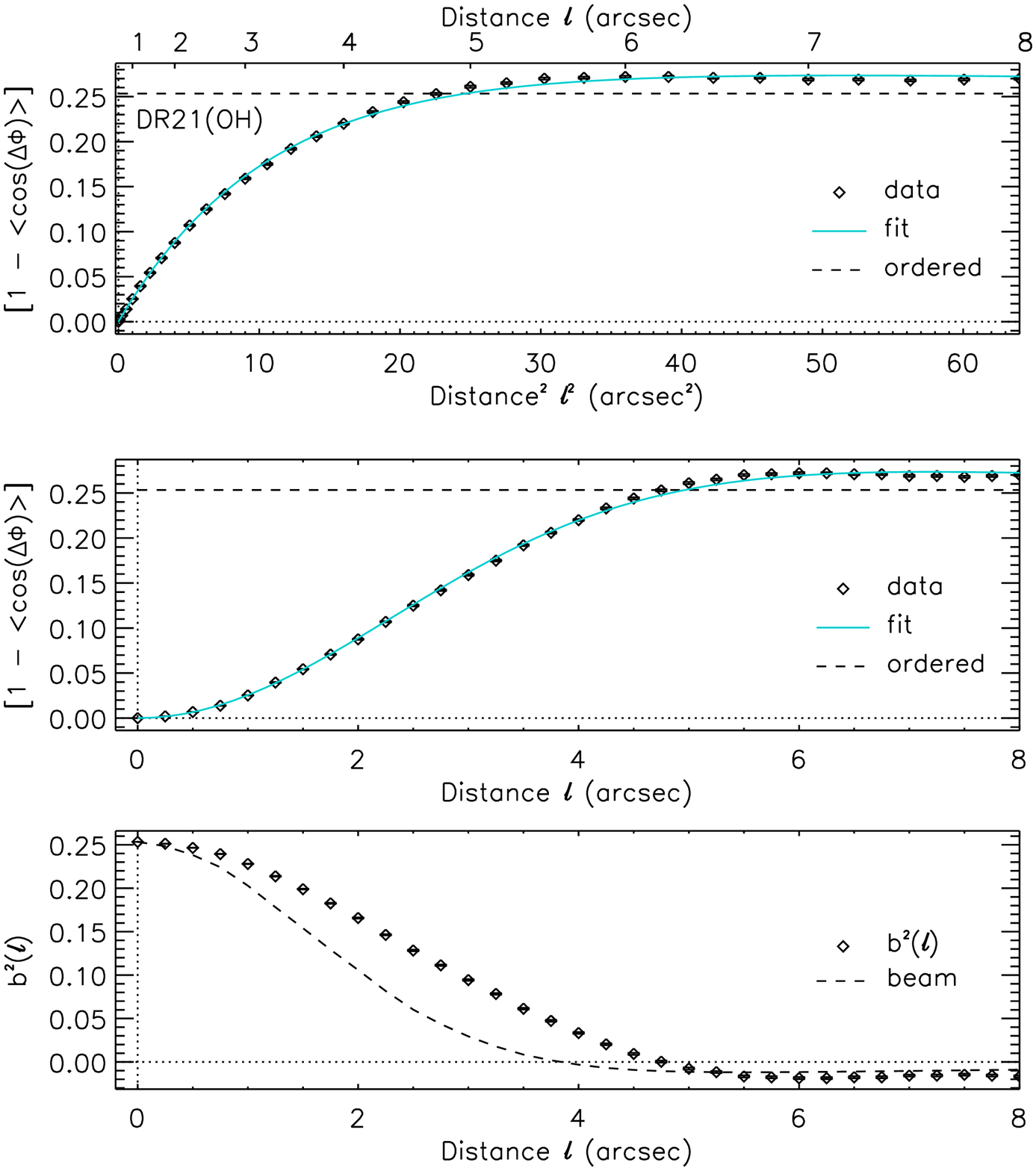}

\caption{\label{fig:DR21(OH)-dispersion}Same as Figure \ref{fig:W3(OH)-dispersion}
but for DR21(OH), where the fit to the data yields $\delta\simeq1\farcs69\pm0\farcs02$
(or $12.3\pm0.2$ mpc at the distance of W3 Main) and $\left\langle B_{\mathrm{t}}^{2}\right\rangle /\left\langle B^{2}\right\rangle \simeq0.70\pm0.01$.}
\end{figure}

\begin{figure}
\plotone{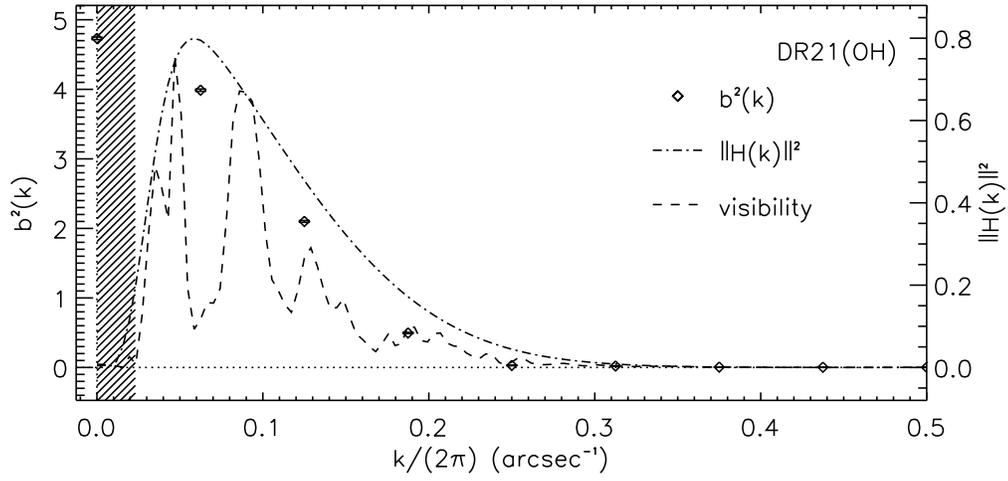}

\caption{\label{fig:DR21(OH)-spectrum}Same as Figure \ref{fig:W3(OH)-spectrum}
but for DR21(OH), where $W_{1}=1\farcs1$ and $W_{2}=7\farcs6$.}
\end{figure}

DR21(OH) is a massive star-forming region located some 1500 pc away.
The CARMA map on which our analysis is performed is centered at $\mathrm{RA}(\mathrm{J2000})=20^{\mathrm{h}}39^{\mathrm{m}}01\fs1$,
$\mathrm{Decl}(\mathrm{J2000})=42^{\circ}22\arcmin29\farcs0$, and
will be found in Figure 32 of \citet{Hull2014}. The FWHM of the synthesized
telescope beam for these observations is $2\farcs7\times2\farcs6$
at a $\mathrm{PA}=37.5^{\circ}$, yielding $W_{1}=1\farcs1$ and $W_{2}=7\farcs6$
for our analysis, while $\Delta^{\prime}=13\farcs5$. The results
of the dispersion analysis for this source are presented in Figure
\ref{fig:DR21(OH)-dispersion} and Figure \ref{fig:DR21(OH)-spectrum}.
As for the other sources studied here, our model yielded a good fit
to the data providing $\delta=1\farcs69\pm0\farcs02$ (or $12.3\pm0.2$
mpc at the distance of DR21(OH)), $\left\langle B_{\mathrm{t}}^{2}\right\rangle /\left\langle B^{2}\right\rangle =0.70\pm0.01$,
and $N=6.91\pm0.07$. 

The total flux measured from the CARMA map is $4.6\:\mathrm{Jy}$,
yielding a total mass of $65\:M_{\sun}$ (we again use a dust temperature
of $30\:\mathrm{K}$ and $\kappa_{\nu}\simeq0.02\;\mathrm{cm^{2}\:g^{-1}}$).
This result is about a factor of 2 smaller than the $150\:M_{\sun}$
estimate obtained by \citet{Girart2013} with previous SMA interferometry
data at $880\:\micron$. We find that this mass is approximately contained
within a circle of $12.4\arcsec$ (FWHM), as seen on the plane of
the sky, giving a mean density of $\simeq2.0\times10^{6}\:\mathrm{cm^{-3}}$
(or $\rho\simeq7.8\times10^{-18}\:\mathrm{g}\,\mathrm{cm}^{-3}$).
We use the velocity dispersion of $1\:\mathrm{km\:s}^{-1}$ from the
$\mathrm{H^{13}CO}{}^{+}\left(4\rightarrow3\right)$ SMA observations
of \citet{Girart2013} and find $B_{0}\sim1.2\:\mathrm{mG}$. Still
precise within a factor of approximately three, this value is consistent
with the Zeeman $\mathrm{CN}\left(1\rightarrow0\right)$ measurement
of \citet{Falgarone2008} who found values of $0.36\:\mathrm{mG}$
and $0.71\:\mathrm{mG}$ for line of sight component of the magnetic
field for the MM1 and MM2 components in this source. The same type
of agreement exists with the earlier result of \citet{Hezareh2010, Hezareh2014}
who obtained $B_{0}\simeq0.7\:\mathrm{mG}$ with single-dish $\mathrm{H^{13}CN}\left(4\rightarrow3\right)$
and $\mathrm{H^{13}CO^{+}}\left(4\rightarrow3\right)$ data using
the ion-neutral line width comparison technique developed by \citet{Li2008b}
(see also \citealt{Houde2000a, Houde2000b, Houde2001}). Furthermore,
our value of $\delta=12.3$ mpc is in good agreement with the 8.5
mpc dissipation scale they measured using the same method, as we would
expect that scale to be shorter than the turbulence correlation length
(see Paper II and III). Finally, it is interesting to note that our
analysis also gives values that are reasonably close to those obtained
by \citet{Girart2013} through their analysis of the aforementioned
SMA interferometry data that yielded $B_{0}\simeq2.1\:\mathrm{mG}$
and $\delta\simeq16.9\:\mathrm{mpc}$, although they used the dispersion
analysis technique developed for single-dish data in Paper II (see
below). 

\begin{deluxetable}{lcccccccc}
\tabletypesize{\scriptsize}
\tablewidth{0pt}
\tablecolumns{11}

\tablehead{
\colhead{Source} & \colhead{$S_\nu$ (Jy)\tablenotemark{a}} & \colhead{$M_\mathrm{gas} \: \left(M_\sun\right)$} & \colhead{$\rho \: \left(\mathrm{g\, cm}^{-3}\right)$} & \colhead{$\sigma\left(v\right) \: \left(\mathrm{km\,s}^{-1}\right)$} & \colhead{$\delta$ (mpc)} & \colhead{$\left<B^2_\mathrm{t}\right>/\left<B^2\right>$} & \colhead{$N$\tablenotemark{b}} & \colhead{$B_0$ (mG)\tablenotemark{c}}
}

\tablecaption{Dispersion Analysis Results.
\label{tab:results}}

\startdata

W3(OH) & $5.7$ & $149$ & $5.0\times10^{-18}$ & $1.1$ & $19.0\pm0.2$ & $0.58\pm0.01$ & $4.67\pm0.04$ & $1.1$ \\
W3 Main & $3.5$ & $83$ & $1.8\times10^{-18}$ & $1.2$ & $22.2\pm0.3$ & $0.74\pm0.01$ & $9.58\pm0.04$ & $0.7$ \\
DR21(OH) & $4.6$ & $65$ & $7.8\times10^{-18}$ & $1.0$ & $12.3\pm0.2$ & $0.70\pm0.01$ & $6.91\pm0.07$ & $1.2$ \\

\enddata

\tablenotetext{a}{From Equation (\ref{eq:M_gas}) with $T_\mathrm{d}=30$ K and $\kappa_\nu=0.02 \: \mathrm{cm}^2\,\mathrm{g}^{-1}$ for all sources.}
\tablenotetext{b}{Number of turbulent cells in the column of gas subtended by the telescope beam.}
\tablenotetext{c}{Accurate within a factor of approximately three.}

\end{deluxetable}

\section{Summary and Conclusion\label{sec:Conclusion}}

Although an idealization that is not likely to be realized in the
ISM, the Gaussian turbulence model provides a useful analytical solution
to the angular dispersion analysis problem, either for single-dish
or interferometry observations, in that it allows the quantification
of key parameters characterizing magnetized turbulence in the ISM.
It is therefore interesting to note that despite this idealization,
our application of the model to interferometry has yielded an excellent
fit for each of the three data sets presented in this paper. 

As noted earlier in Section \ref{sub:interferometer}, this new solution
for interferometry differs from the one for single-dish data by a
dependency on $W_{2}$, the width of the Gaussian function subtracted
to the synthesized interferometry beam (of width $W_{1})$ to account
for the filtering of extended structures (i.e., low spatial frequencies).
Since this interferometry solution for the angular dispersion function
(i.e., Equation {[}\ref{eq:int_solution}{]}) tends to the single-dish
solution (i.e., Equation {[}\ref{eq:SD-solution}{]}) in the limit
when $W_{1}\ll W_{2}$, it is at this point interesting to assess
the errors that would ensue if the single-dish model was used for
the analysis of our interferometry data. More precisely, for the three
sources, all with good $uv$-coverage at low frequencies, we have
$0.1\lesssim W_{1}/W_{2}\lesssim0.2$ and find that the errors we
incur are relatively modest. That is, the single-dish model overestimates
the values for $b^{2}\left(0\right)$ and the intrinsic turbulent-to-ordered
magnetic energy ratio $\left\langle B_{\mathrm{t}}^{2}\right\rangle /\left\langle B_{0}^{2}\right\rangle $
by approximately $10\%$ to $20\%$, while the turbulence correlation
length $\delta$ and the number of turbulent cells $N$ (or $N_{1}$
for the single-dish) are underestimated by $\sim5\%$. As was earlier
alluded to, it is likely that the CLEAN algorithm used in processing
the interferometry data could inject some signals at low-frequencies
in the power spectrum \citep{Thompson2004}. For such cases, our twin-Gaussian
beam model probably overestimates the filtering effect from the interferometer
beam. Still, the single- and twin-Gaussian beam models provide results
for two opposite limiting cases (i.e, for the single-dish with minimum
filtering at low-frequencies and the interferometer with maximum filtering)
and allow to specify a range for the dispersion analysis output parameters.
For the present cases, the significant uncertainties on some of the
other parameters entering the estimates obtained for magnetic field
strengths with the DCF equation (e.g., $\rho$ or $\kappa_{\nu}$),
any error on the output parameters would have a small contribution
to the overall uncertainty on any magnetic field strength estimate.
But this is probably more a statement on the difficulties encountered
when trying to indirectly evaluate magnetic field strengths with techniques
not relying on the Zeeman effect. On the other hand, with the future
availability of high spatial resolution polarization ALMA data with
excellent $uv$-coverage, it is likely that errors on the order of
$10\%$ will become more important when finely characterizing magnetized
turbulence in a similar manner as was presented here or through its
power spectrum (as in Paper III). 

For the TADPOL/CARMA data of \citet{Hull2014} presented in this paper,
our analysis yielded a turbulence correlation length $\delta\simeq19$
mpc, a ratio of turbulent-to-total magnetic energy $\left\langle B_{\mathrm{t}}^{2}\right\rangle /\left\langle B^{2}\right\rangle \simeq0.58$,
and magnetic field strength $B_{0}\sim1.1\:\mathrm{mG}$ for W3(OH);
$\delta\simeq22$ mpc, $\left\langle B_{\mathrm{t}}^{2}\right\rangle /\left\langle B^{2}\right\rangle \simeq0.74$,
and $B_{0}\sim0.7\:\mathrm{mG}$ for W3 Main; and $\delta\simeq12$
mpc, $\left\langle B_{\mathrm{t}}^{2}\right\rangle /\left\langle B^{2}\right\rangle \simeq0.70$,
and $B_{0}\sim1.2\:\mathrm{mG}$ for DR21(OH). These three sources
therefore appear to have a significant amount of magnetic energy in
the form of turbulence. Finally, our estimates for the turbulence
correlation length and magnetic field strengths are consistent with
corresponding values obtained from other sources, sometimes obtained
with different techniques.

\acknowledgements{M.H.'s research is funded through the NSERC Discovery Grant, Canada
Research Chair, Canada Foundation for Innovation, Ontario Innovation
Trust, and Western's Academic Development Fund programs. CARMA development
and operations were funded by the National Science Foundation and
the CARMA partner universities.}

\appendix
\section{Data Analysis}\label{sec:Appendix}

Given the angle difference between a pair of data points separated
by $\ell_{ij}\equiv\vert\mathbf{r}_{i}-\mathbf{r}_{j}\vert$

\begin{equation}
\Delta\Phi_{ij}=\Phi_{i}-\Phi_{j}\label{eq:diff_ij}
\end{equation}

\noindent we calculate the mean $\left\langle \cos\left(\Delta\Phi_{ij}\right)\right\rangle _{k}$
from the data for $\left(\ell_{k}-\Delta\ell/2\right)\leq\ell_{ij}<\left(\ell_{k}+\Delta\ell/2\right)$,
with $\ell_{k}=k\Delta\ell$ an integer multiple of the grid spacing
$\Delta\ell=0\farcs25$. This function is then corrected for measurement
uncertainties according to

\begin{equation}
\left\langle \cos\left(\Delta\Phi_{ij}\right)\right\rangle _{k,0}\simeq\frac{\left\langle \cos\left(\Delta\Phi_{ij}\right)\right\rangle _{k}}{1-\frac{1}{2}\left\langle \sigma^{2}(\Delta\Phi_{ij})\right\rangle _{k}},\label{eq:cos_ij_corr}
\end{equation}

\noindent where the uncertainty on $\Delta\Phi_{ij}$ is given by

\begin{equation}
\sigma^{2}(\Delta\Phi_{ij})\simeq\sigma^{2}(\Phi_{i})+\sigma^{2}(\Phi_{j})-2\sigma(\Phi_{i})\sigma(\Phi_{j})e^{-\ell_{ij}^{2}/4W_{1}^{2}}\label{eq:sigma2_dphi}
\end{equation}

\noindent and $\sigma^{2}(\Phi_{i})$ is the uncertainty on $\Phi_{i}$.
Equation (\ref{eq:sigma2_dphi}) thus takes into account that pairs
of data points will be correlated when separated by approximately
less than the telescope beam. From this, the measurement uncertainties
for the dispersion function $1-\left\langle \cos\left(\Delta\Phi_{ij}\right)\right\rangle _{k,0}$
are determined through 

\begin{eqnarray}
\sigma^{2}\left[\left\langle \cos\left(\Delta\Phi_{ij}\right)\right\rangle _{k,0}\right] & = & \left\langle \sin\left(\Delta\Phi_{ij}\right)\right\rangle _{k}^{2}\left[\left\langle \sigma^{2}(\Delta\Phi_{ij})\right\rangle _{k}+\left\langle (\Delta\Phi_{ij})^{2}\right\rangle _{k}\right]\nonumber \\
 &  & +\left[\frac{3}{4}\left\langle \cos\left(\Delta\Phi_{ij}\right)\right\rangle _{k}^{2}-\left\langle \sin\left(\Delta\Phi_{ij}\right)\right\rangle _{k}^{2}\right]\left[\left\langle \sigma^{2}(\Delta\Phi_{ij})\right\rangle _{k}+\left\langle (\Delta\Phi_{ij})^{2}\right\rangle _{k}\right]^{2},\label{eq:sigma2(cos)}
\end{eqnarray}

\noindent for all $\left(\ell_{k}-\Delta\ell/2\right)\leq\ell_{ij}<\left(\ell_{k}+\Delta\ell/2\right)$.
Although this analysis follows similar presentations found in \citet{Houde2009, Houde2013},
Equation (\ref{eq:sigma2(cos)}) was augmented to better account for
the different sources of uncertainty.

\end{document}